\documentclass[10pt,twocolumn]{wlscirep}
\usepackage{graphicx}
\usepackage{amsmath,amssymb}
\usepackage{booktabs,longtable}
\usepackage{tikz}
\usepackage{listings}
\usepackage{tcolorbox}
\usepackage{xcolor}
\tcbuselibrary{listings, breakable}
\usepackage{placeins}
\usepackage{wrapfig}
\usepackage[version=4]{mhchem}
\usepackage{chemformula} 
\mhchemoptions{text-greek=default}
\usepackage{siunitx}
\usepackage{textcomp}
\usepackage[colorlinks=true,allcolors=blue]{hyperref}

\usepackage{caption}
\usepackage{ragged2e}
\AtBeginDocument{\justifying}

\title{AI-Driven Structure
Refinement of X-ray Diffraction}

\author[1]{Bin Cao}
\author[2,4]{Qian Zhang}
\author[2]{Zhenjie Feng}
\author[1]{Taolue Zhang}
\author[1]{Jiaqiang Huang}
\author[1,3]{Lu-Tao Weng}
\author[2,1,$*$]{Tong-Yi Zhang}

\affil[1]{Guangzhou Municipal Key Laboratory of Materials Informatics, Advanced Materials Thrust, Sustainable Energy and Environment Thrust, The Hong Kong University of Science and Technology (Guangzhou), Guangzhou, 511453, China}

\affil[2]{Materials Genome Institute, Shanghai University, Shanghai 200444, China}

\affil[3]{Material Characterization and Preparation Facility, Hong Kong University
of Science and Technology (Guangzhou), Guangzhou 511400, China}

\affil[4]{Hefei National Research Center for Physical Sciences at the Microscale, and Hefei National Laboratory, University of Science and Technology of China, Hefei, Anhui 230026, 
China}

\affil[$*$]{Corresponding authors: mezhangt@hkust-gz.edu.cn}

\date{}

\begin{document}
\maketitle
\begin{abstract}

Artificial intelligence can rapidly propose candidate phases and structures from X-ray diffraction (XRD), but these hypotheses often fail in downstream refinement because peak intensities cannot be stably assigned under severe overlap and diffraction consistency is enforced only weakly. Here we introduce the whole-pattern expectation--maximization (WPEM) algorithm, a physics-constrained whole-pattern decomposition and refinement workflow that turns Bragg's law into an explicit constraint within a batch expectation--maximization framework. WPEM models the full profile as a probabilistic mixture density and iteratively infers component-resolved intensities while keeping peak centres Bragg-consistent, producing a continuous, physically admissible intensity representation that remains stable in heavily overlapped regions and in the presence of mixed radiation or multiple phases. 
We benchmark WPEM on standard reference patterns (PbSO$_4$ and Tb$_2$BaCoO$_5$), where it yields lower $R_p/R_{wp}$ than widely used packages (FullProf and TOPAS) under matched refinement conditions. We further demonstrate generality across realistic experimental scenarios, including phase-resolved decomposition in multiphase materials, quantitative recovery of mixture compositions, separation of crystalline peaks from amorphous backgrounds in semicrystalline systems, high-throughput operando lattice tracking, automated refinement of compositionally disordered solid solutions, and quantitative phase-resolved analysis of complex archaeological samples from synchrotron powder XRD. 
By providing Bragg-consistent, uncertainty-aware intensity partitioning as a refinement-ready interface, WPEM closes the gap between AI-generated hypotheses and diffraction-admissible structure refinement on challenging XRD data.

\end{abstract}

\makeatletter
\ifx\theabstract\undefined\else
  {\absfont\textbf{Abstract}\par\theabstract\par}\vskip10pt
\fi
\makeatother

\keywords{}

\section*{Main}

Powder diffraction is approaching its centenary, yet the field is being reshaped by a distinctly contemporary force: artificial intelligence (AI). X-ray diffraction (XRD) remains the most widely deployed non-destructive probe of atomic structure, enabling quantitative access to crystal symmetry, lattice distortions, phase composition, defects and residual stress from routine measurements \cite{dinnebier2015powder,bunaciu2015x}. In parallel, AI is rapidly changing how materials are discovered and understood, from generative proposal of candidate crystal structures \cite{zeni2025generative,jiao2024space,antunes2024crystal,zhu2024wycryst,park2025exploration,cheng2026artificial}, to fast atomistic relaxation via machine-learning force fields \cite{yang2024mattersim,gao2025foundation,chen2022universal,wang2025martini3}, and even direct inference of structure from powder XRD (PXRD) data \cite{li2025powder,davel2025machine,szymanski2023autonomous,lai2025end}. Together, these advances suggest a realistic, high-throughput vision: end-to-end pipelines that convert diffraction measurements into physically valid atomic models with minimal human intervention.

\begin{figure*}[t]
    \centering
    \includegraphics[width=\textwidth,height=0.95\textheight,keepaspectratio]{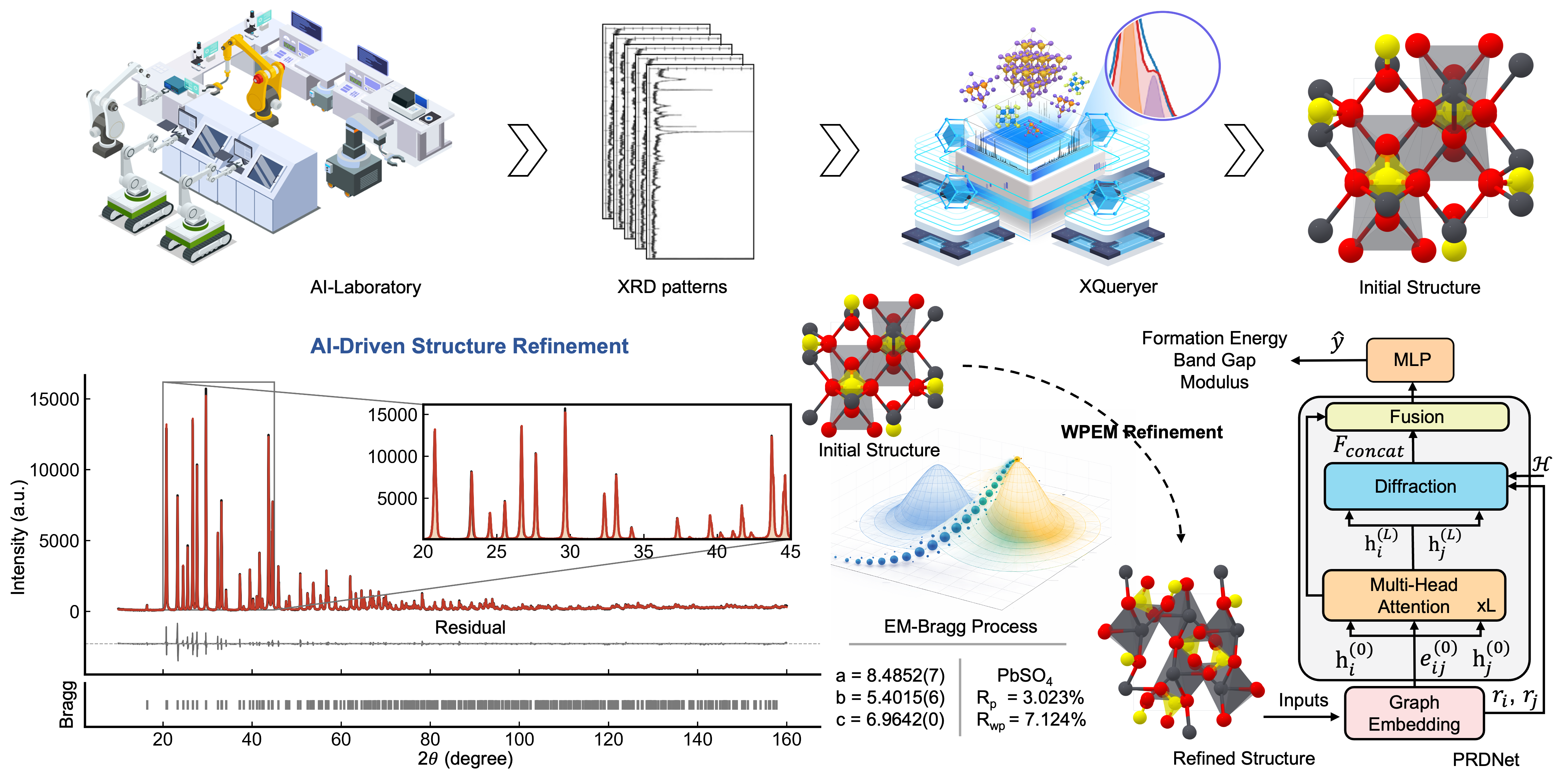}
    \caption{\textbf{From AI-generated hypotheses to physically admissible refined structures.} Given a batch of experimental diffraction patterns, the AI phase/structure identification model Xqueryer\cite{cao2025xqueryer} proposes an initial structural hypothesis from the raw PXRD data. The subsequent EM--Bragg refinement performs Bragg-law-constrained whole-pattern decomposition to enforce diffraction consistency and yield a physically grounded refined structure. Finally, the property-prediction model PRDNet\cite{cao2025beyond} estimates key materials properties (e.g., formation energy, band gap and elastic moduli) to support downstream screening and optimization.}
    \label{fig:TOP}
\end{figure*}

Realizing this vision, however, hinges on a bottleneck that current AI workflows rarely resolve: enforcing physical and crystallographic consistency during refinement. AI-generated or AI-deciphered structures may violate basic constraints (symmetry, chemistry, stability, or diffraction consistency), and such failures are increasingly visible as implausible structures propagate into major databases \cite{Chawla2025DuplicateStructures}. For PXRD in particular, the inverse problem is ill-posed: the one-dimensional profile compresses three-dimensional reciprocal-space information, and severe peak overlap can make intensity partitioning ambiguous even when lattice parameters are approximately known. Consequently, existing AI-based workflows can become unstable, over-parameterized, or drift toward solutions that fit the data numerically while violating diffraction physics.
Importantly, this ill-posedness is not only algorithmic but also physical. The measured XRD profile is jointly governed by instrument factors (e.g., source spectrum, optics, divergence, zero shift, detector response and resolution) and sample factors (e.g., crystal structure, lattice strain, size broadening, preferred orientation, texture and phase mixture). High-quality measurements and well-characterized instrument functions sharpen peak shapes and improve peak separation, thereby strengthening the reliability of any downstream analysis.

AI-driven crystallography must therefore be built around a physics-anchored refinement layer that explicitly couples machine inference to diffraction theory. In this framework, AI is best used to propose hypotheses (phases, lattice metrics and candidate structures), whereas refinement must enforce the hard constraints that make those hypotheses experimentally admissible. Concretely, we view AI-enabled PXRD analysis as a three-stage process: (1) PXRD-based phase/structure identification (Xqueryer\cite{cao2025xqueryer}) to anchor inference to plausible structural priors \cite{bin2025simxrd}; (2) whole-pattern decomposition under explicit Bragg-law constraints to resolve overlapped reflections into diffraction-consistent peak parameters; and (3) structure--property inference (PRDNet \cite{cao2025beyond}) to connect diffraction signatures to functional metrics (Fig.~\ref{fig:TOP}).

Here we introduce whole-pattern expectation–maximization (WPEM) algorithm, a physics-constrained whole-pattern decomposition and refinement workflow designed for a common XRD failure mode: once reflections overlap, intensities become ill-determined and refinement can drift away from diffraction-consistent peak assignments. WPEM models the profile as a probabilistic mixture and infers component-resolved intensities with a batch expectation--maximization (EM)\cite{do2008expectation} algorithm under an explicit Bragg-law constraint, keeping peak centres Bragg-consistent while separating overlap through responsibility weighting. It outputs a continuous, uncertainty-aware intensity representation that remains stable in congested regions and is robust to mixed radiation and multiphase data.
WPEM is meant to complement, not replace, AI-based phase/structure identification: it acts as a physics-anchored refinement layer that converts AI-generated hypotheses into diffraction-admissible inputs for subsequent crystallographic refinement by providing Bragg-consistent peak partitions with fewer effective degrees of freedom.

\subsection*{Diffraction-consistent probabilistic decomposition}

Bragg peak positions, intensities and line shapes jointly encode geometric and chemical information, but extracting these quantities from a one-dimensional powder profile is intrinsically ill-posed because multiple reflections can coincide in $2\theta$ \cite{dinnebier2015powder,debye1916interferenzen}. Classical whole-pattern refinement methods (Rietveld, Pawley and Le Bail) enforce crystallographic consistency by coupling all reflections through a single structural model \cite{rietveld1967line,rietveld1969profile,pawley1981unit,le1988ab}. Under severe overlap, however, the resulting least-squares problem can be numerically fragile and may admit non-physical solutions (e.g., poorly determined or even negative intensities).

WPEM addresses this bottleneck by recasting whole-pattern decomposition as probabilistic inference subject to hard diffraction constraints. Specifically, we treat the background-subtracted profile as a mixture of peak-shape functions (PSFs) and estimate component responsibilities via EM, which naturally handles strong overlap through soft assignments. The observed intensity profile $y(2\theta)$ is approximated as
\begin{equation}
 y(2\theta) \approx \sum_{i=1}^{m} w_i\, p_i(2\theta;\,\boldsymbol{\phi}_i),
 \label{eq:mixture_main}
\end{equation}
where $p_i$ is a normalized PSF (we use a pseudo-Voigt family in practice), $w_i \ge 0$ is the component weight (proportional to integrated intensity), and $\boldsymbol{\phi}_i$ collects peak-shape parameters.

In WPEM, this mixture view is made explicit by interpreting the background-subtracted diffraction profile as a continuous \emph{measure function} assembled from $m$ peak-shape functions. Each PSF is modelled using a thin-tailed pseudo-Voigt (PV) density, i.e., a convex combination of Lorentzian and Gaussian components,
\begin{equation}
\small
p_{\mathrm{pv}}(x)=
\Delta\cdot\frac{1}{\pi}\frac{\gamma}{(x-\mu)^2+\gamma^2}
+ (1-\Delta)\cdot\frac{1}{\sqrt{2\pi}\sigma}\exp\!\left(-\frac{(x-\mu)^2}{2\sigma^2}\right),
\end{equation}
where $\mu$ is the Bragg-peak centre, $\gamma$ is the Lorentzian half width at half maximum (HWHM), $\sigma$ is the Gaussian standard deviation, and $\Delta\in(0,1)$ controls the Lorentzian--Gaussian mixing (with $\Delta\to1$ recovering a Lorentzian and $\Delta\to0$ recovering a Gaussian). Unlike Le Bail-type profile matching, which often constrains the Lorentzian and Gaussian widths through a shared full width at half maximum (FWHM) model (for example, $\Gamma^2=U\tan^2\theta+V\tan\theta+W$), we treat $\gamma$ and $\sigma$ as independent variables and optimize them directly within EM--Bragg to accommodate heterogeneous broadening and severe overlap.

The closed-form update rules in the $(t+1)$-th maximization step are given by
\begin{align}
\mu_i^{(t+1)}
&=
\frac{
\frac{2\pi}{\gamma_i^{(t)}}
\sum_{j=1}^{N}
y_j^{\mathrm{nb}}
\beta_{ji}^{L(t)} x_j
+
\frac{1}{\sigma_i^{2(t)}}
\sum_{j=1}^{N}
y_j^{\mathrm{nb}}
T_{ji}^{R(t)} x_j
}{
\frac{2\pi}{\gamma_i^{(t)}}
\sum_{j=1}^{N}
y_j^{\mathrm{nb}}
\beta_{ji}^{L(t)}
+
\frac{1}{\sigma_i^{2(t)}}
\sum_{j=1}^{N}
y_j^{\mathrm{nb}}
T_{ji}^{R(t)}
},
\label{8a}\\
\gamma_i^{(t+1)}
&=
\left[
\frac{
\sum_{j=1}^{N}
y_j^{\mathrm{nb}}
\beta_{ji}^{L(t)}
\big(x_j-\mu_i^{(t)}\big)^2
}{
\sum_{j=1}^{N}
y_j^{\mathrm{nb}}
\beta_{ji}^{L(t)}
}
\right]^{1/2},
\label{8b}\\
\sigma_i^{2(t+1)}
&=
\frac{
\sum_{j=1}^{N}
y_j^{\mathrm{nb}}
T_{ji}^{R(t)}
\big(x_j-\mu_i^{(t)}\big)^2
}{
\sum_{j=1}^{N}
y_j^{\mathrm{nb}}
T_{ji}^{R(t)}
},
\label{8c}\\
\Delta_i^{(t+1)}
&=
\frac{
\sum_{j=1}^{N}
y_j^{\mathrm{nb}}
T_{ji}^{L(t)}
}{
\sum_{j=1}^{N}
y_j^{\mathrm{nb}}
T_{ji}^{(t)}
}.
\label{8d}
\end{align}

Here $y_j^{\mathrm{nb}}$ denotes the background-subtracted intensity at grid point $j$, $T_{ji}^{L}$ is the (normalized) responsibility/weight that assigns point $j$ to peak component $i$ in the E-step (with superscripts indicating the update used), and $\beta_{ji}^{L}$ and $T_{ji}$ are auxiliary quantities used in the closed-form M-step updates; their explicit definitions are given in the Supporting Information.

Under the Bragg-law constraint on the peak centres, Eqs.~(\ref{8a})--(\ref{8d}) are iterated to update the mixture parameters, including the peak position $\mu_i$, the Lorentzian and Gaussian width parameters $\gamma_i$ and $\sigma_i^2$, and the mixing coefficient $\Delta_i$. Full details of the statistical model and the derivation of the EM--Bragg procedure are provided in the Supporting Information.

Early, in-house deployments of the WPEM concept in our group have supported several ongoing structure-determination studies \cite{qin2023orthorhombic,cao2024active,lei2024li,shi2025dissecting,zhang2024crystallographic,qin2025ferromagnetic,sun2024machine}. These efforts reflect a key perspective: structure determination is intrinsically multimodal, and reliable conclusions require integrating complementary constraints (diffraction, imaging, spectroscopy, and theory/simulation). In this context, WPEM acts as a bridging layer by providing diffraction-consistent intensity partitioning from PXRD that can be directly combined with other probes. For example, in our VS$_4$ cathode study \cite{shi2025first}, WPEM and transmission electron microscopy provided cross-validated constraints that enabled confident identification of the operando transformation structure (CCDC 2487046) \cite{CCDC2487046}. Here, after several years of continued development, we present WPEM as a reproducible, unified framework and evaluate it systematically across challenging scenarios: reference-pattern benchmarking under severe overlap (\ce{PbSO4}, \ce{Tb2BaCoO5}); phase-resolved decomposition in a multiphase Ti--15Nb thin film; composition recovery for \ce{NaCl}--\ce{Li2CO3} mixtures; high-throughput operando lattice tracking; automated refinement of a disordered Ru--Mn oxide solid solution; separation of crystalline peaks from amorphous halos in semicrystalline polymers; and deciphering a 3,000-year-old recipe from synchrotron PXRD of an ancient Egyptian make-up sample.

\subsection*{Benchmarking and quantitative phase analysis}

\begin{figure*}[!h]
    \centering
    \includegraphics[width=\textwidth,height=0.95\textheight,keepaspectratio]{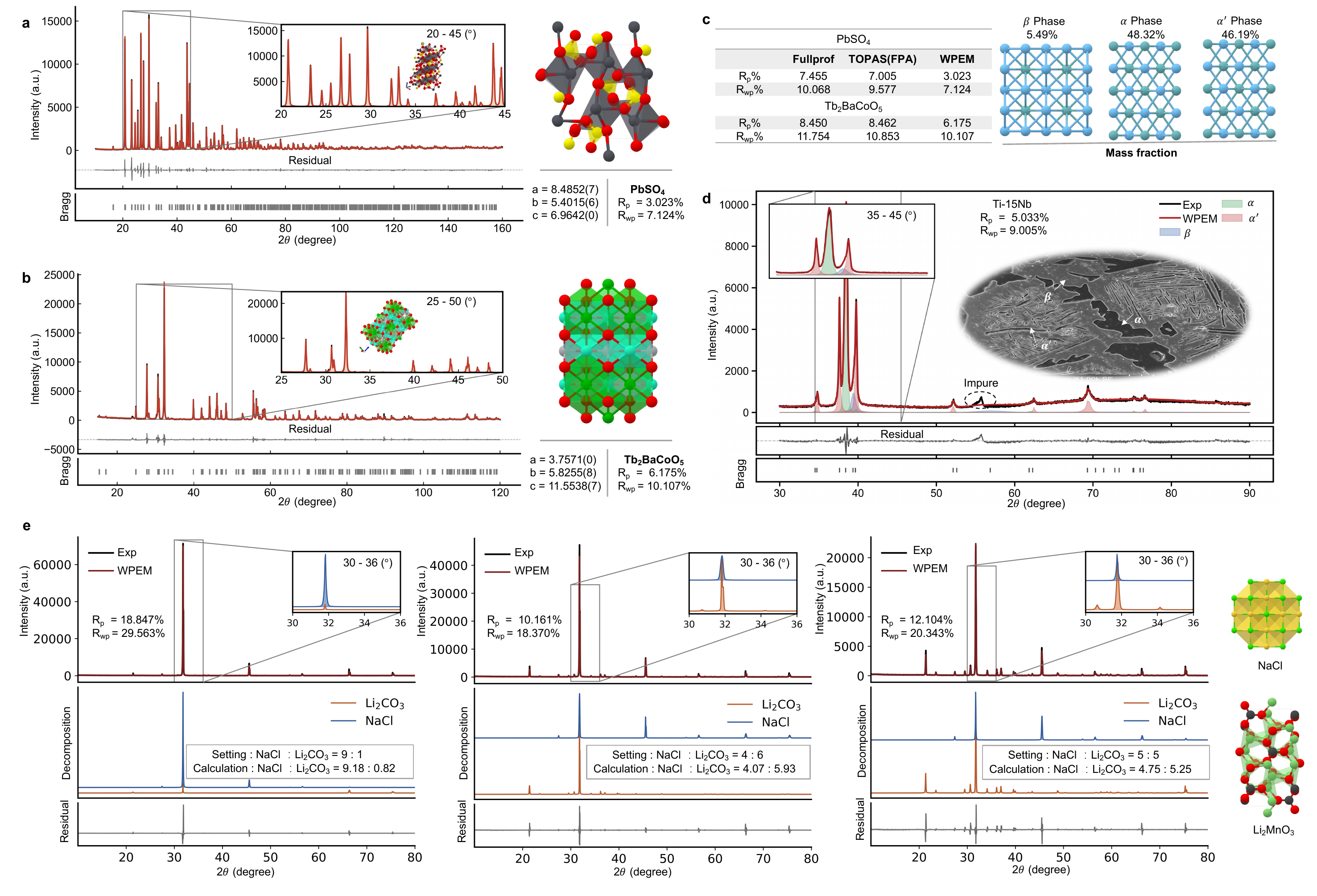}
    \caption{\textbf{Benchmarking and quantitative decomposition.} \textbf{a,b} Whole-pattern decomposition of reference data (\ce{PbSO4} and \ce{Tb2BaCoO5}) and comparison of agreement factors against FullProf and TOPAS. \textbf{c} Summary of profile factors for WPEM, FullProf and TOPAS (FPA). \textbf{d} Decomposition of a multiphase Ti--15Nb thin film into $\beta$,$\alpha$ and $\alpha'$ contributions and the corresponding phase fractions estimated from integrated intensities. \textbf{e} Quantitative composition analysis of \ce{NaCl}--\ce{Li2CO3} powder mixtures showing agreement between inferred and nominal mass ratios.}
    \label{fig:Benchmark}
\end{figure*}

To ensure reproducibility and enable quantitative comparisons, we report the profile factors ($R_{\mathrm{p}}$, $R_{\mathrm{wp}}$; lower is better) \cite{toby2006r} together with the number of free profile parameters, the $2\theta$ range, and the radiation wavelengths used in each refinement. For all methods, backgrounds are treated consistently (polynomial/segmented) and refined on the same background-subtracted profiles to avoid confounding differences in baseline modelling. Unless stated otherwise, peak positions are constrained by a common set of lattice parameters for each phase at initialization, and scale factors are refined globally.

Beyond standard $R$-factors, we assess robustness under overlap by comparing refinements across representative datasets with strong peak overlap and by inspecting the stability of the fitted profiles under consistent refinement settings. WPEM is less sensitive to local overlap because EM responsibilities provide a soft, overlap-aware partition of intensity, whereas least-squares profile matching can become ill-conditioned when multiple reflections fall within a single instrumental resolution element.

We first benchmarked WPEM against two widely used whole-pattern refinement packages with complementary modeling philosophies: FullProf profile matching (Le Bail fit with a constant scale factor) \cite{rodriguez2001fullprof} TOPAS (fundamental-parameters approach, FPA) \cite{dinnebier2018rietveld}. The head-to-head comparison was carried out on two standard reference datasets distributed with these packages, \ce{PbSO4} (Pnma) and \ce{Tb2BaCoO5} (Immm) (Fig.~\ref{fig:Benchmark}\textbf{a,b}). These two cases represent a demanding regime in which peak overlap is pervasive and refinement stability becomes critical.

For \ce{PbSO4}, which contains 383 pairs of Bragg reflections under a Cu source ($K_{\alpha 1}=1.540560\,\text{\AA}$; $K_{\alpha 2}=1.544330\,\text{\AA}$), WPEM achieves $R_{\mathrm{p}}=3.023\%$ and $R_{\mathrm{wp}}=7.124\%$ (Fig.~\ref{fig:Benchmark}\textbf{a}). The refined lattice parameters are $a=8.4852(7)\,\text{\AA}$, $b=5.4015(6)\,\text{\AA}$, $c=6.9642(0)\,\text{\AA}$ and $\alpha=\beta=\gamma=90^\circ$. For \ce{Tb2BaCoO5} (131 reflection pairs), WPEM yields $R_{\mathrm{p}}=6.175\%$ and $R_{\mathrm{wp}}=10.107\%$ (Fig.~\ref{fig:Benchmark}\textbf{b}), with refined lattice parameters $a=3.7571(0)\,\text{\AA}$, $b=5.8255(8)\,\text{\AA}$, $c=11.5538(7)\,\text{\AA}$ and $\alpha=\beta=\gamma=90^\circ$.

Across both benchmarks, WPEM reduces profile factors relative to FullProf and TOPAS, with the largest gains observed in regions dominated by heavy peak overlap (Fig.~\ref{fig:Benchmark}\textbf{c}). We attribute this improved stability to the combination of soft assignment in the EM updates and the Bragg-consistent constraints that prevent peak centers from drifting into non-physical solutions. The full fitted profiles and refinement settings for FullProf and TOPAS are provided in the Supplementary Information.

We next demonstrate a capability that is essential for AI-driven materials analysis: resolving closely related phases with subtly different lattice metrics. For a Ti--15Nb (atomic ratio) alloy thin film containing $\beta$ (bcc), $\alpha$ (hcp) and $\alpha^{\prime}$ (hcp) phases (Fig.~\ref{fig:Benchmark}\textbf{d}), distinguishing $\alpha$ and $\alpha^{\prime}$ is challenging because the martensitic $\alpha^{\prime}$ phase forms during rapid cooling and differs only slightly from $\alpha$, leading to strongly overlapped reflections. WPEM separates the phase-resolved diffraction contributions with $R_{\mathrm{p}}=5.033\%$ and $R_{\mathrm{wp}}=9.005\%$. The refined lattice parameters are: $\beta$ phase $a=b=c=3.2345(9)\,\text{\AA}$, $\alpha=\beta=\gamma=90^\circ$; $\alpha$ phase $a=b=2.9977(8)\,\text{\AA}$, $c=4.6817(3)\,\text{\AA}$, $\alpha=\beta=90^\circ$, $\gamma=120^\circ$; and $\alpha^{\prime}$ phase $a=b=2.9785(4)\,\text{\AA}$, $c=4.7768(3)\,\text{\AA}$, $\alpha=\beta=90^\circ$, $\gamma=120^\circ$. From the separated component profiles, we estimate phase fractions of $\beta\!:\!\alpha\!:\!\alpha^{\prime} = 5.49\%\!:\!48.32\%\!:\!46.19\%$.

Finally, we evaluated quantitative composition analysis on a controlled, standards-based mixture test, using binary powder mixtures of \ce{NaCl} and \ce{Li2CO3} prepared at nominal mass ratios of 9:1, 4:6 and 5:5. This benchmark is deliberately stringent for whole-pattern methods: the two phases exhibit multiple overlapping reflections and markedly different peak intensities, so small biases in intensity partitioning can translate into large errors in inferred composition. Using the phase-resolved component profiles returned by WPEM, we estimate NaCl mass fractions of 91.8\% (nominal 90\%), 40.7\% (nominal 40\%) and 47.5\% (nominal 50\%) (Fig.~\ref{fig:Benchmark}\textbf{e}). The close agreement across three composition regimes, NaCl-rich, \ce{Li2CO3}-rich and near-equimolar, supports that WPEM not only improves profile fits but also yields quantitatively reliable decompositions suitable for mixture analysis.

\subsection*{Operando tracking and solid-solution refinement}

\begin{figure}[h]
    \centering
    \includegraphics[width=\linewidth]{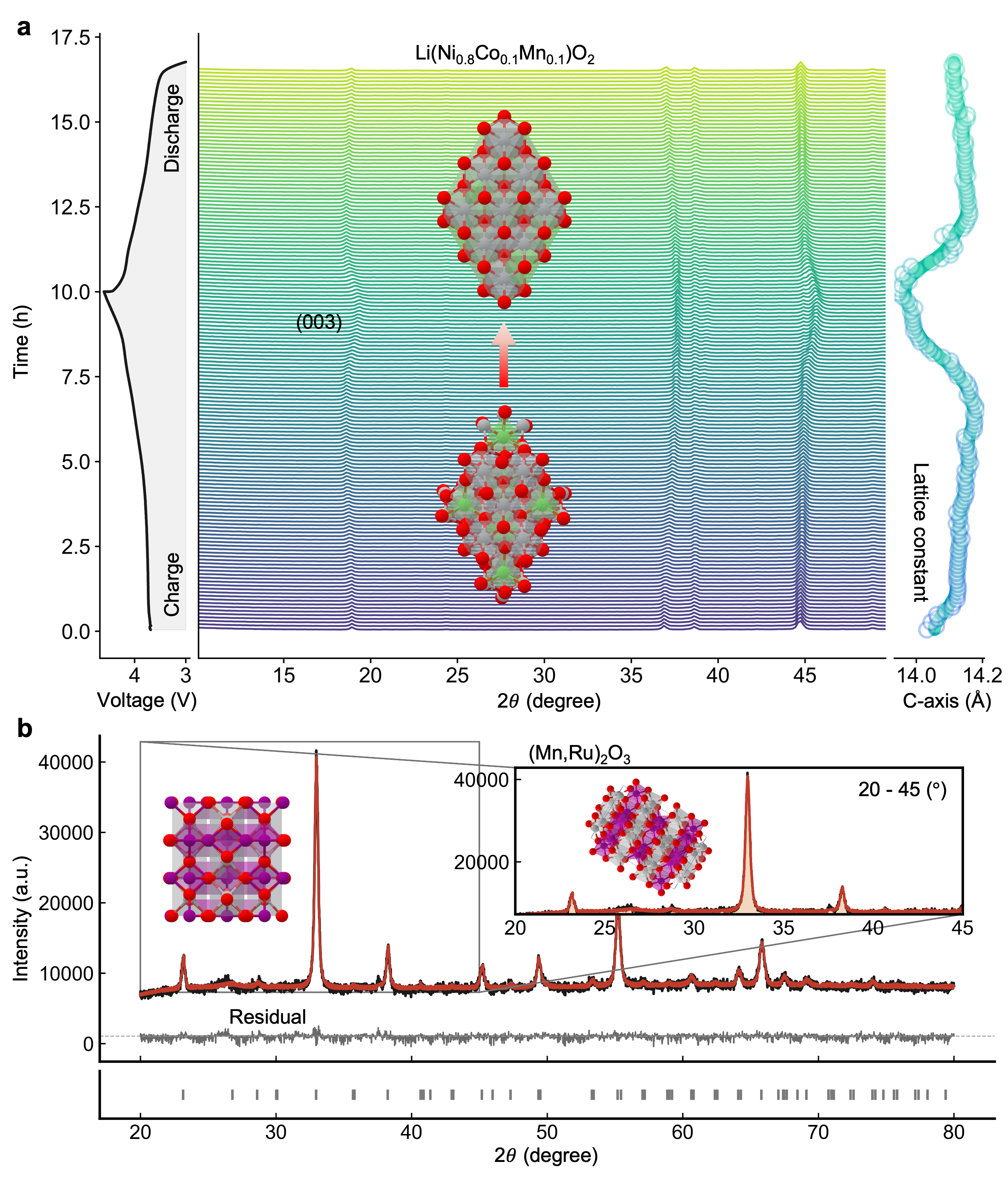}
     \caption{\textbf{High-throughput PXRD workflows enabled by WPEM.} \textbf{a} In~situ tracking of lattice-parameter evolution in $\mathrm{Li(Ni_{0.8}Co_{0.1}Mn_{0.1})O_2}$ during electrochemical cycling. \textbf{b} Automated solid-solution refinement of a Ru-substituted Mn oxide using a $3\times3$ supercell model (CCDC~2530452).}

    \label{fig:throughput}
\end{figure}

WPEM is well suited to operando XRD, where large time-series datasets demand robust, automated whole-pattern refinement. Here we analyzed operando XRD patterns of an O3-type $\mathrm{Li(Ni_{0.8}Co_{0.1}Mn_{0.1})O_2}$ cathode collected during electrochemical cycling between 3.0 and 4.6~V. Using batch WPEM, we refined each pattern and tracked the evolution of the lattice parameter $c$ during charge and discharge. In this workflow, the refined lattice parameters from one scan initialize the next, stabilizing convergence and enabling consistent lattice tracking across the full series without manual intervention. Importantly, the background of each pattern is refined independently, without a background template or a predefined set of refinement parameters. To keep the procedure reproducible and strictly data-driven, we did not impose explicit cross-pattern constraints (e.g., forcing neighboring scans to share lattice constants); thus, the extracted trajectories are determined by the diffraction data of each scan. The refined lattice parameters show that the $c$ axis first expands and then collapses sharply at high states of charge, consistent with the peak shifts in the raw operando profiles. By converting operando ``peak motion'' into refined lattice constants through whole-pattern fitting, WPEM provides an objective, refinement-ready route to quantify lattice evolution throughout the measurement (Fig.~\ref{fig:throughput}\textbf{a}).

We further demonstrate WPEM on a Ru--Mn oxide electrocatalyst synthesized via cation exchange, starting from orthorhombic $\mathrm{Mn_2O_3}$ prepared by a hydrothermal route and reacted with aqueous $\mathrm{RuCl_3}$ solutions \cite{qin2023orthorhombic}. By varying the $\mathrm{RuCl_3}$ dosage, the Ru loading in the final products can be tuned. Because catalytic stability and activity are sensitive to the local Ru configuration, resolving the resulting solid-solution structure is essential for mechanistic understanding. Here, WPEM was used to screen candidate Ru site occupations in a $3\times3$ supercell model and identify a configuration that best matches the experimental pattern, achieving $R_{\mathrm{p}}=1.447\%$ and $R_{\mathrm{wp}}=3.129\%$ (Fig.~\ref{fig:throughput}\textbf{b}). The resolved structure has been deposited at the Cambridge crystallographic data centre (CCDC) under deposition number 2530452 \cite{CCDC2530452}. This example highlights how WPEM can automate structure refinement in complex, compositionally disordered systems while maintaining diffraction consistency.

\subsection*{Deciphering ancient Egyptian make-up by phase-resolved decomposition}

\begin{figure*}[!h]
    \centering
    \includegraphics[width=\textwidth,height=0.9\textheight,keepaspectratio]{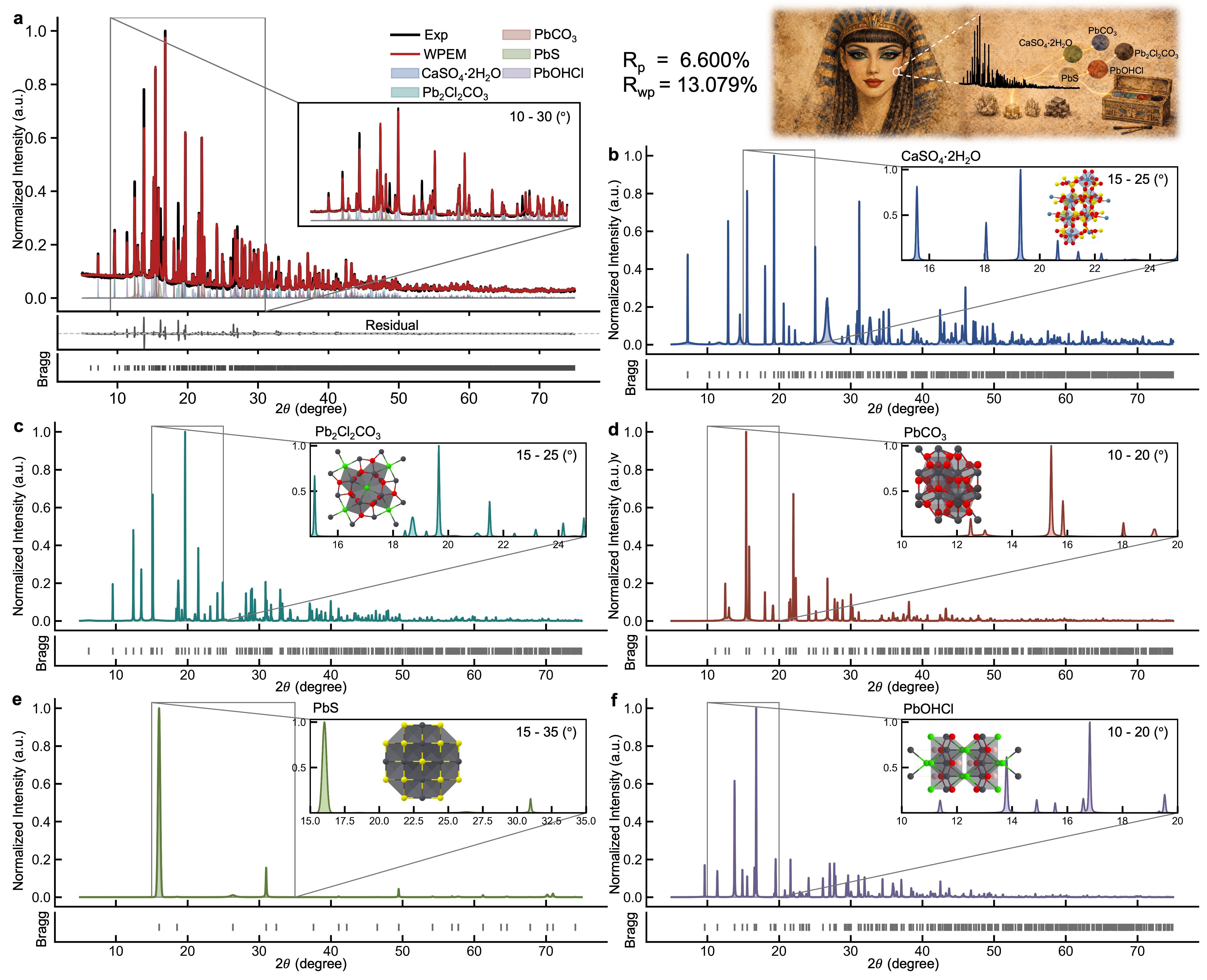}
    \caption{\textbf{Deciphering an ancient Egyptian make-up sample.} \textbf{a,} Synchrotron PXRD profile (points) and WPEM fit (line) collected with $\lambda = 0.96270\,\text{\AA}$. \textbf{b--f,} Phase-resolved component patterns recovered by WPEM for \ce{CaSO4.2H2O} (gypsum), \ce{Pb2Cl2CO3} (phosgenite), \ce{PbCO3} (cerussite), \ce{PbS} (galena) and \ce{PbOHCl} (laurionite), respectively. The decomposition quantitatively separates strongly overlapped reflections in this complex multiphase mixture, enabling phase-level interpretation directly from the full profile. The Egyptian portrait was generated by ChatGPT-5o and does not infringe any third-party copyrights.}
    \label{fig:makeup}
\end{figure*}

Mineral-based make-up was used extensively in ancient Egypt and has been documented since the earliest periods of Egyptian history \cite{forbes1965studies}. A particularly compelling dataset is the exceptionally well-preserved cosmetic powders dated to 2000--1200~BC analysed by Walter \textit{et al.} using non-destructive synchrotron diffraction and complementary chemical analysis \cite{walter1999making}. Their work identified a characteristic, strongly overlapped multiphase assemblage comprising gypsum (\ce{CaSO4.2H2O}), galena (\ce{PbS}), cerussite (\ce{PbCO3}), and the rare lead chloride phases laurionite (\ce{PbOHCl}) and phosgenite (\ce{Pb2Cl2CO3}) \cite{walter1999making}. The presence of laurionite and phosgenite, phases uncommon in natural deposits and often associated with corrosion products, was interpreted as evidence of deliberate wet-chemical synthesis rather than simple in-container weathering \cite{walter1999making}.

This system provides an attractive ``stress test'' for whole-pattern decomposition: multiple phases contribute dense forests of peaks, and intense overlap makes manual, reflection-by-reflection assignment impractical. As shown in Fig.~\ref{fig:makeup}, WPEM decomposes the full synchrotron PXRD profile into phase-resolved component patterns for gypsum, phosgenite, cerussite, galena and laurionite. From the component intensities (see Supplementary Information), we estimate mass fractions of 12.53\%, 18.53\%, 32.02\%, 9.69\% and 27.23\% for these phases, respectively, providing a clear recipe for     \textbf{reconstructing this 3{,}000-year-old formulation}. Beyond producing an accurate fit, the phase-resolved decomposition yields a quantitative, interpretable inventory of ingredients from a single, highly overlapped diffraction profile, illustrating how physics-constrained decomposition can support data-driven crystallographic ``forensics'' on real experimental samples.

The refined lattice parameters (in \AA{} and degrees) are: gypsum $a=5.6800(9)$, $b=15.2144(0)$, $c=6.5303(7)$, $\alpha=\gamma=90^\circ$, $\beta=118.4841(5)^\circ$; phosgenite $a=b=8.1600(0)$, $c=8.8834(3)$, $\alpha=\beta=\gamma=90^\circ$; cerussite $a=5.1794(7)$, $b=8.4922(9)$, $c=6.1418(1)$, $\alpha=\beta=\gamma=90^\circ$; galena $a=b=c=5.9388(0)$, $\alpha=\beta=\gamma=90^\circ$; and laurionite $a=9.7002(6)$, $b=4.0200(3)$, $c=7.1108(1)$, $\alpha=\beta=\gamma=90^\circ$.

\begin{figure}[h!]
    \centering
    \includegraphics[width=1\columnwidth]{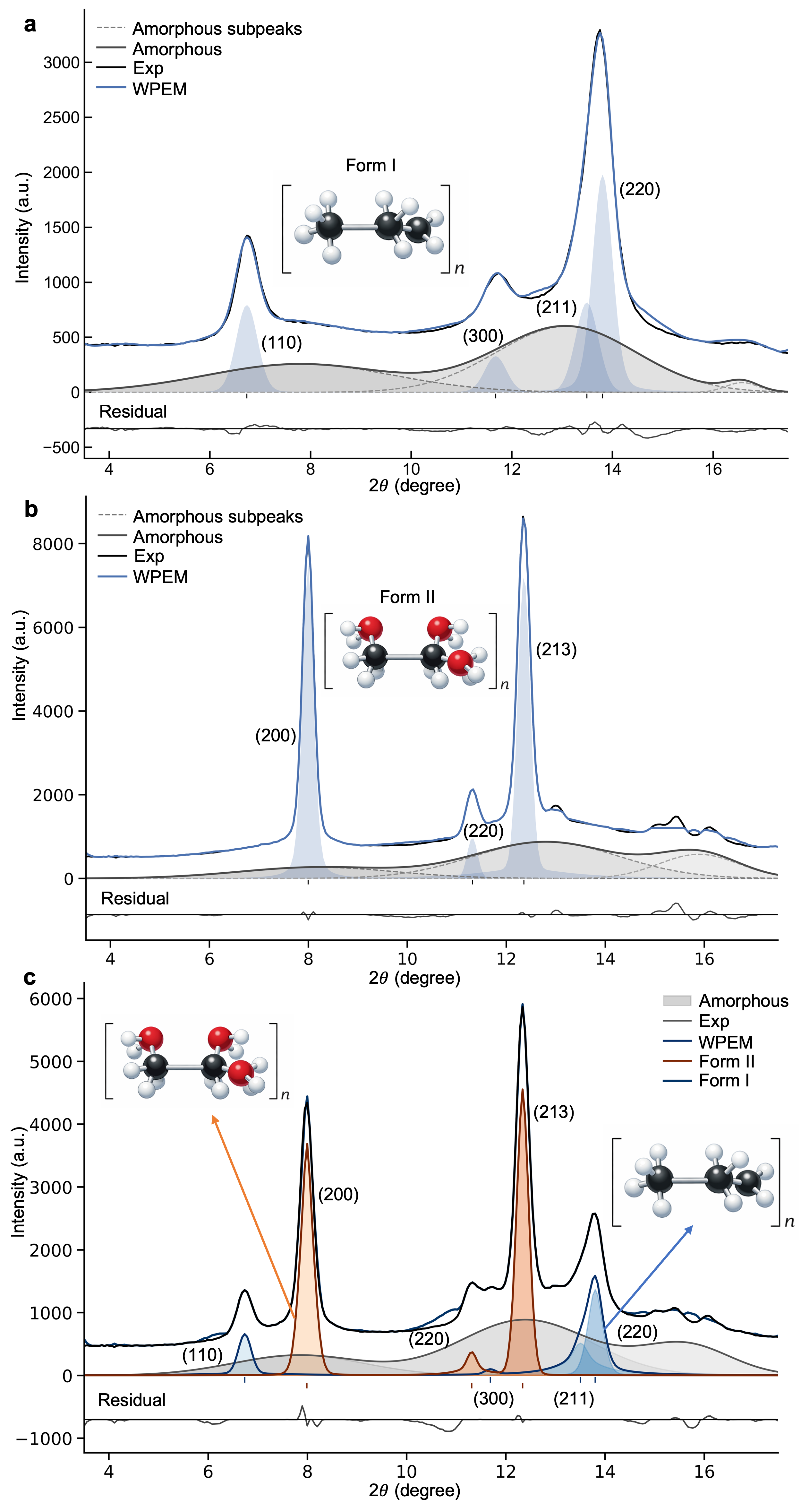}
     \caption{\textbf{Quantitative analysis of semicrystalline diffraction.}  \textbf{a} Decomposition of the PE form~I pattern into four crystalline peaks and three amorphous halos. \textbf{b} Decomposition of the PE form~II pattern into three crystalline peaks and three amorphous halos. \textbf{c} Decomposition of the PB-I diffraction profile into form~I, form~II and amorphous-halo contributions.}
    
    \label{fig:amorphous}
\end{figure}

\subsection*{Semicrystalline decomposition under strain: separating crystalline peaks and amorphous halos}

Semicrystalline polymers pose a distinctive challenge for whole-pattern decomposition: sharp Bragg reflections from crystalline lamellae coexist with broad amorphous scattering, and both evolve continuously under deformation. To demonstrate that WPEM can quantify such coupled crystalline--amorphous dynamics, we analyse \textit{in situ} wide-angle X-ray diffraction (WAXD) data from polybutene-I (PB-I) reported by Feng \textit{et al.} \cite{feng2022strain}, who investigated a strain-rate-dependent phase transition mechanism during uniaxial stretching using ultrafast synchrotron WAXD. PB-I exhibits multiple polymorphs (forms~I, I$'$, II and III), and the mechanically accelerated transformation from metastable form~II to stable form~I is a canonical yet overlap-dominated test case.

Figures~\ref{fig:amorphous}\textbf{a,b} first illustrate the key capability of WPEM on semicrystalline patterns (PB-I form I and form II): it separates multiple crystalline peaks from broad amorphous halos in a single, unified fit, avoiding ad hoc background subtraction. Figure~\ref{fig:amorphous}\textbf{c} then applies the same framework to PB-I, decomposing the measured profile into three physically interpretable components, PB form~I, PB form~II and amorphous halos, while preserving Bragg-consistent peak positions for the crystalline phases. The recomposed profile reproduces the experimental pattern with $R_{\mathrm{p}}=1.725\%$ and $R_{\mathrm{wp}}=4.315\%$.

Crucially, the PB-I profile contains severe peak overlap: the form~I reflections (200) and (211) occur at nearly the same scattering angle and are difficult to resolve without additional constraints. WPEM mitigates this ambiguity through overlap-aware responsibilities, yielding a stable component decomposition in which these reflections can be distinguished in the recovered patterns. In addition, the simultaneous separation of crystalline peaks and amorphous halos enables direct quantification of bulk crystallinity from the integrated component areas,
\[
\chi_c=\frac{\sum A_c}{\sum A_c+\sum A_a}\times100\%=45.11\%,
\]
where $A_c$ and $A_a$ denote the integrated areas of crystalline peaks and amorphous halos, respectively. This example highlights an important advantage of WPEM for polymer diffraction: it provides a physically interpretable, quantitative partition of intensity between overlapping crystalline polymorphs and the amorphous phase within a single whole-pattern framework.

\section*{Conclusion}
AI models are increasingly capable of proposing candidate phases and structures from PXRD, but turning these hypotheses into scientifically admissible results still requires a refinement layer that enforces diffraction physics. Here we present WPEM as such a ``physics anchor'': by casting whole-pattern decomposition as probabilistic inference and solving it with a Bragg-law-constrained EM scheme, WPEM delivers diffraction-consistent peak partitions even under severe overlap, mixed radiation and multiphase conditions.

This physics-constrained decomposition layer is complementary to data-driven models and naturally fits multimodal pipelines. Upstream, it can stabilize AI-enabled phase/structure identification by converting raw profiles into Bragg-consistent, component-resolved representations that can be fused with complementary evidence (e.g., microscopy, spectroscopy, and prior chemical/thermodynamic constraints). Downstream, it provides physically meaningful, uncertainty-aware inputs for structure refinement and property prediction, reducing the risk of overfitting and non-physical solutions. More broadly, integrating WPEM into end-to-end, multimodal AI workflows offers a practical route from ``pattern-to-hypothesis'' to ``pattern-to-structure'' that is scalable, automated and grounded in diffraction theory.

\bibliography{references}

\section*{Data availability}
All datasets used in this study, together with the corresponding scripts and results required to reproduce the analyses, are publicly available (see \url{https://github.com/Bin-Cao/PyWPEM/tree/main/CASES}).

\section*{Code availability}
WPEM is open source at \url{https://github.com/Bin-Cao/PyWPEM}. Documentation and reproducible examples are available at \url{https://pyxplore.netlify.app/}.

\section*{Supplementary Information}
The Supplementary Information contains detailed derivations of our method and is available at \url{https://doi.org/10.6084/m9.figshare.31374217}.

\section*{Acknowledgements}
This work was supported by the Guangzhou--HKUST(GZ) Joint Funding Program (2023A03J0003 and 2023A03J0103). We also acknowledge Prof.~Ren-Chao Che (Fudan University), Prof.~Wei Chen (University of Science and Technology of China), Prof.~Peng Ding (Shanghai University), Dr.~Yin Qin (Harbin Institute of Technology, Shenzhen), Dr.~Xiu-Ling Shi (Harbin Institute of Technology, Shenzhen), Dr.~Xiao-Yun Guo (Shanghai University), and Dr.~Yu Huang (Shanghai University) for sharing data from their published work and for helpful discussions.

\section*{Author Contribution}
 Bin Cao led the study, carried out the core methodological development, data analysis, and drafted the manuscript. Qian Zhang contributed to code development for single-phase fitting. Zhenjie Feng performed the FullProf and TOPAS refinements and obtained the Egyptian powder XRD dataset. Taolue Zhang prepared and measured the standard \ce{NaCl}--\ce{Li2CO3} mixture samples. Jiaqiang Huang purchased the standard \ce{NaCl}--\ce{Li2CO3} samples and contributed to discussion and interpretation. Lu-Tao Weng contributed to experimental coordination and provided critical input on data interpretation and manuscript revision. Tong-Yi Zhang conceived the project, supervised the research, and provided critical guidance on diffraction methodology and manuscript revision.

\section*{Competing interests}
The authors declare no competing interests.

\end{document}